# Possible Coexistence of Antihydrogen with Hydrogen, Deuterium and Tritium Atoms

**Mohamed Assad Abdel-Raouf (*)**

**Department of Physics, College of Science, UAEU,**

**Al-Ain, P.O. Box: 17555, United Arab Emirates**

## Abstract

Recent productions of large numbers of cold antiprotons as well as the formation of antihydrogens at CERN and Fermilab have raised basic questions about possible coexistence of matter and antimatter in nature. In the present work, previous mathematical considerations are revisited which support the possible coexistence of Antihydrogen with Hydrogen, Deuterium and Tritium atoms. In particular, the main objective of the present work is to present computational treatments which confirm the possible formation of these quasi molecules in laboratory. These treatments are based on a nonadiabatic picture of the system in which generalized basis functions are adjusted within the framework of Rayleigh-Ritz' variational method. Thus, it is ruled out in the present work the Born-Oppenheimer adiabatic picture of the system, which demands the existence of bound states composed of fixed quasi heavy atoms (containing at least two baryons, e.g. protonium (Pn), with mean lifetime $\tau \sim 1.0x10^{-6}$ s) and quasi light atoms (composed of two leptons, e.g. positronium (Ps), with $\tau = 125x10^{-12}$ s for para-Ps and $\tau = 142.05x10^{-9}$ s for ortho-Ps). Our calculations of the binding energies and internal structure of Antihydrogen-Hydrogen, Antihydrogen-Deuterium and Antihydrogen-Tritium show that these quasi molecules are bound and could be formed in nature. On the other hand, having in mind the adiabatic picture of the systems, our results suggest the possible formation of these molecules as resonant states in Antihydrogen-Atom interaction. Nevertheless, several arguments are accumulated in the conclusion as consequences of the proposed bound states.

---

**(*) E-mail:** assad@uaeu.ac.ae

# 1. INTRODUCTION

Although the world of particle physics was discovered in the early thirties of the preceding century with Anderson's pioneering work on the anti-electrons (positrons) and the justification of Dirac's theory of holes, the great realization of the field was accomplished in the fifties after the discovery of the antiprotons and their production in laboratories. The crucial developments in this direction were made in USA and Europe through the construction of circular high energy accelerators at Fermi and CERN laboratories. The energy of the accelerated particles has been increased over the years. It started at CERN in 1957 with a proton synchrotron of energy 600 MeV and should reach this year the order of 2-3 tetra eV with the establishment of the large hadron colloider (LHC). Consequently the number of antiparticles produced in Labs should increase drastically. Another important step in the world of particle physics was achieved through the discovery of new cooling techniques at CERN in 1982. It then became possible to produce antiprotons, cool and accumulate them at few MeV. This led to the establishment of the famous Low Energy Antiproton Ring (LEAR). Later development of more advanced Antiproton Traps has contributed drastically to the production of the first anti-atom in Lab, namely the Antihydrogen through the "ATHENA" experiment at CERN in 1995 [1] and the confirmation made in 1997 at Fermi's Lab [2]. Very recent improvements upon the antiproton traps in the two Labs allowed in 2002 for the development of highly populated Antihydrogen traps [3], [4]. Consequently, previous theoretical discussions [5]-[7] were reopened regarding the possible formation of Antihydrogen-Hydrogen exotic molecules at laboratories.

As a matter of fact the Antihydrogen-Hydrogen (or more general Antimatter-Matter) problem has been computationally investigated in the literatures on two levels. On the first level, it was treated as a collision process in which scattering parameters (e.g. scattering lengths [8]) and cross sections [9]-[12] are calculated. In this case the quasi molecular structures may show up as resonant states, the annihilation rates of which are determined (see e.g. [13]-[17]). On the second level, the problem was tackled as a bound state problem subjected to the Born-Oppenheimer picture of the four-body system. The interest was concentrated on searching for possible existence of a potential barrier which prevents the dissociation into protoniums and positroniums and, therefore, supports the formation of the quasi molecules. Thus, the main attention was directed to the possible formation of a bound state composed of a positronium (Ps) and fixed protonium (Pn) quasi atoms. Particularly, most of the interest was devoted to the calculation of the smallest internal distance of the protonium, below which the two quasi atoms are unbound. The value of this distance was decreasing over the years leading to quite contradicted conclusions. Whilst many authors (see e.g. [18]-[20]), have emphasized that such a bound state could not exist, recent results (see e.g. [21]-[24]) encouraged the opinion that these conclusions are not decisive, (for a review, see [25]).

On the other hand, a theoretical proof was established two decades ago [26-29], for the possible coexistence of four-body systems of the form $A^-B^+B^-A^+$ at arbitrary values of the mass ratio between negatively and positively charged particles, i.e. $\sigma = m_A/M_B$, running between 0 and $\infty$. It was demonstrated that the binding energy $W(\sigma)$ of the system is a continuous concave function of $\sigma$ lying inside the triangle (0, W(0)), (0,

W(1)), (1, W(1)) and W(1) is an upper bound for all W(σ), with σ falling between 0 and ∞. Three points should be noticed here:

(i) The binding energy is, in contrast with the dissociation energy, the amount of energy which keeps the atoms $A^-B^+$ bound in an exotic system.

(ii) The case σ = 0 corresponds to the Born-Oppenheimer treatment of the hydrogen (or antihydrogen) molecule, while σ = 1 refers to a system composed of two identical particles and their two identical antiparticles, e.g. $e^- e^+ e^+ e^-$, $p\bar{p}\,\bar{p}p$, i.e. the so called positronium ( $Ps_2$) [30] and protonium ($Pn_2$) molecules.

(iii) There is no critical value of σ at which the molecule should not exist.

Thus, the theorem and its conclusions suggested that the systems, e.g.$e^-\pi^+\pi^-e^+$, $e^-\mu^+\mu^-e^+$, $\mu^-\pi^+\pi^-\mu^+$, should exist as molecular structures. Elaborate computational calculations [31] have confirmed the theorem and its conclusions as well as the lastly mentioned point.

Consequently, the possible formation of an exotic four-body molecule composed of the constituents of the antihydrogen and hydrogen atoms is confirmed on the basis of a rigorous mathematical argument [32]. The theorem was also extended [33] to prove the possible formation of antihydrogen-deuterium and antihydrogen-tritium molecules. In all previously mentioned works it was ruled out the Born-Oppenheimer treatment of the four-body molecules where two light particles should be moving in the field of fixed two heavy particles.

The aim of the present work is to investigate the possible coexistence of antihydrogens with hydrogen, deuterium and tritium atoms. (The corresponding quasi molecules will be frequently referred to as Heterohydrogens). This should be accomplished by calculating their binding energies and internal distances using Rayleigh-Ritz variational method (RRVM) [34] and the virial theorem (VT) [28]. It is considered that the binding energy is the energy required for preventing dissociation of a molecule into its original constituents. Possible coexistence of subclusters composed of two heavy nuclei (e.g. protoniums) and two light particles (e.g. positroniums) are not precisely considered in the present work. Finally, the consequences of the possible coexistence of matter and antimatter are shortly discussed.

## 2. RAYLEIGH-RITZ VARIATIONAL METHOD

In RRVM [34], a trial wavefunction $\left|\psi_t^{(n)}\right\rangle$ is selected such that

$$|\psi_t^{(n)} = \sum_k^n a_k \left|\psi_{tk}\right\rangle \quad , \tag{1}$$

where n is the dimension of $\left|\psi_{tk}^{(n)}\right\rangle$, and

$$\left\langle\psi_{tk}^{(n)}\middle|\psi_{tk'}^{(n)}\right\rangle = \delta_{kk'} \,, \qquad \text{for } k, k' = 1, 2, \ldots, n \;, \tag{2}$$

where $\delta_{kk'}$ is the Kroneker-delta.

Defining H as the total Hamiltonian, E as the total energy, $\left|\psi_{tk}^{(n)}\right\rangle$'s can be generated from one basis set of vectors $\{\left|\chi_i\right\rangle\} \subset D_H$ , where $D_H$ is the domain of H, i.e.,

$$\left|\psi_{tk}^{(n)}\right\rangle = \sum_{i=1}^{n} c_{ik}\left|\chi_i\right\rangle \quad . \qquad (3)$$

RRVM is subjected to the solution of the following secular equations

$$\sum_{j=1}^{n} c_{jk}\left[\left\langle\chi_i\left|H\right|\chi_j\right\rangle - E_{nk}\left\langle\chi_i\left|\chi_j\right.\right\rangle\right] = 0\,, \quad i = 1,2,\ldots,n\,, \qquad (4)$$

which is meaningful if and only if the determinant $\Delta_{nk}$ satisfies the relation

$$\Delta_{nk} = \det(H_{ij} - E_{nk}S_{ij}) = 0 \quad , \qquad (5)$$

where

$$H_{ij} = \left\langle\chi_i\left|H\right|\chi_j\right\rangle \quad , \qquad (6a)$$

and

$$S_{ij} = \left\langle\chi_i\left|\chi_j\right.\right\rangle \quad . \qquad (6b)$$

The eigenvalues obtained by (6) are ordered such that;

$$E_{n1} \le E_{n2} \le \ldots \le E_{nn} \quad , \qquad (7)$$

is satisfied.

RRVM proves the important relation between $E_{n1}$ and the first (lowest) exact energy level of the system $E_1$, namely that

$$E_1 \le E_{n1}\,, \qquad \text{for } n > 0 \quad , \qquad (8a)$$

i.e., for any choice of the components $\left|\psi_{tk}^{(n)}\right\rangle$, the first variational energy is an upper bound to the exact one. Moreover, if condition (2) is fulfilled, we then get

$$E_k \le E_{nk}\ , \qquad \text{for } k = 1, 2, \ldots n \quad . \qquad (8b)$$

Furthermore, if the trial wavefunction is enlarged by exactly one component, we are left with

$$\left\langle \psi_{tk}^{(n+1)} \middle| \psi_{tk'}^{(n+1)} \right\rangle = \delta_{kk'}\ , \quad \text{for } k, k' = 1, 2, \ldots, n+1 \quad , \qquad (9)$$

Thus, if one or k of $E_{nk}$'s, $k < n$, lie in the negative spectrum of H, in this case, we are unable to make any decisive statement about all states higher than $E_k$ , while the existence of any negative $E_{nk}$'s ensures the existence of all corresponding bound states of the four-body molecule.

The virial theory (VT) , on the other hand, predicts the upper bound $E_k^v$ to the real bound state energy $E_k$ by replacing the coordinates $r_{ij}$ by $\alpha\, r_{ij}$ where $\alpha$ is a variational parameter. Since the kinetic energy operator is second order $(\frac{\partial}{\partial r_{ij}})$ while the potential energy operator is first order of $(\frac{1}{r_{ij}})$, we may define the virial Hamiltonian as

$$H_v = \frac{1}{\alpha^2} T + \frac{1}{\alpha} V \quad , \qquad (10)$$

and since the virial energy $E_k$ is

$$E_k^v = \frac{\left\langle \psi_{tk}^{(n)} \middle| H \middle| \psi_{tk}^{(n)} \right\rangle}{\left\langle \psi_{tk}^{(n)} \middle| \psi_{tk}^{(n)} \right\rangle} \quad ,$$

we have

$$E_k^v = \frac{\left\langle \psi_{tk}^{(n)} \middle| \frac{1}{\alpha^2} T + \frac{1}{\alpha} V \middle| \psi_{tk}^{(n)} \right\rangle}{\left\langle \psi_{tk}^{(n)} \middle| \psi_{tk}^{(n)} \right\rangle} = \frac{1}{\alpha^2} T_{ex} + \frac{1}{\alpha} V_{ex} \quad , \qquad (11)$$

where;

$$T_{ex} = \frac{\left\langle \psi_{tk}^{(n)} \middle| T \middle| \psi_{tk}^{(n)} \right\rangle}{\left\langle \psi_{tk}^{(n)} \middle| \psi_{tk}^{(n)} \right\rangle} \qquad \text{and} \qquad V_{ex} = \frac{\left\langle \psi_{tk}^{(n)} \middle| V \middle| \psi_{tk}^{(n)} \right\rangle}{\left\langle \psi_{tk}^{(n)} \middle| \psi_{tk}^{(n)} \right\rangle} \ . \qquad (12)$$

Now, the upper bound states are obtained by minimization with respect to $\alpha$. Doing so, eq. (11) gives;

$$\frac{\partial E_k^v}{\partial \alpha} = -\frac{2}{\alpha^3} T_{ex} - \frac{1}{\alpha^2} V_{ex} = 0$$

$$\therefore \qquad \alpha = -\frac{2T_{ex}}{V_{ex}} \quad . \qquad (13)$$

Substitute from (12) into (10), we obtain

$$E_k^v = \frac{-V_{ex}^2}{4T_{ex} \left\langle \psi_{tk}^{(n)} \middle| \psi_{tk}^{(n)} \right\rangle} = \frac{-\left\langle \psi_{tk}^{(n)} \middle| V \middle| \psi_{tk}^{(n)} \right\rangle^2}{4 \left\langle \psi_{tk}^{(n)} \middle| T \middle| \psi_{tk}^{(n)} \right\rangle \left\langle \psi_{tk}^{(n)} \middle| \psi_{tk}^{(n)} \right\rangle} \quad , \qquad (14)$$

which is known as the virial energy.

## 3. $\overline{H}H$, $\overline{H}D$ and $\overline{H}T$, EXOTIC MOLECULES

**Fig. (1)** shows a general schematic diagram for the considered four-body molecules, where Mp denotes the mass of the proton (antiproton) and n` Mp (n` ≥ 1) is the mass of the nucleus.

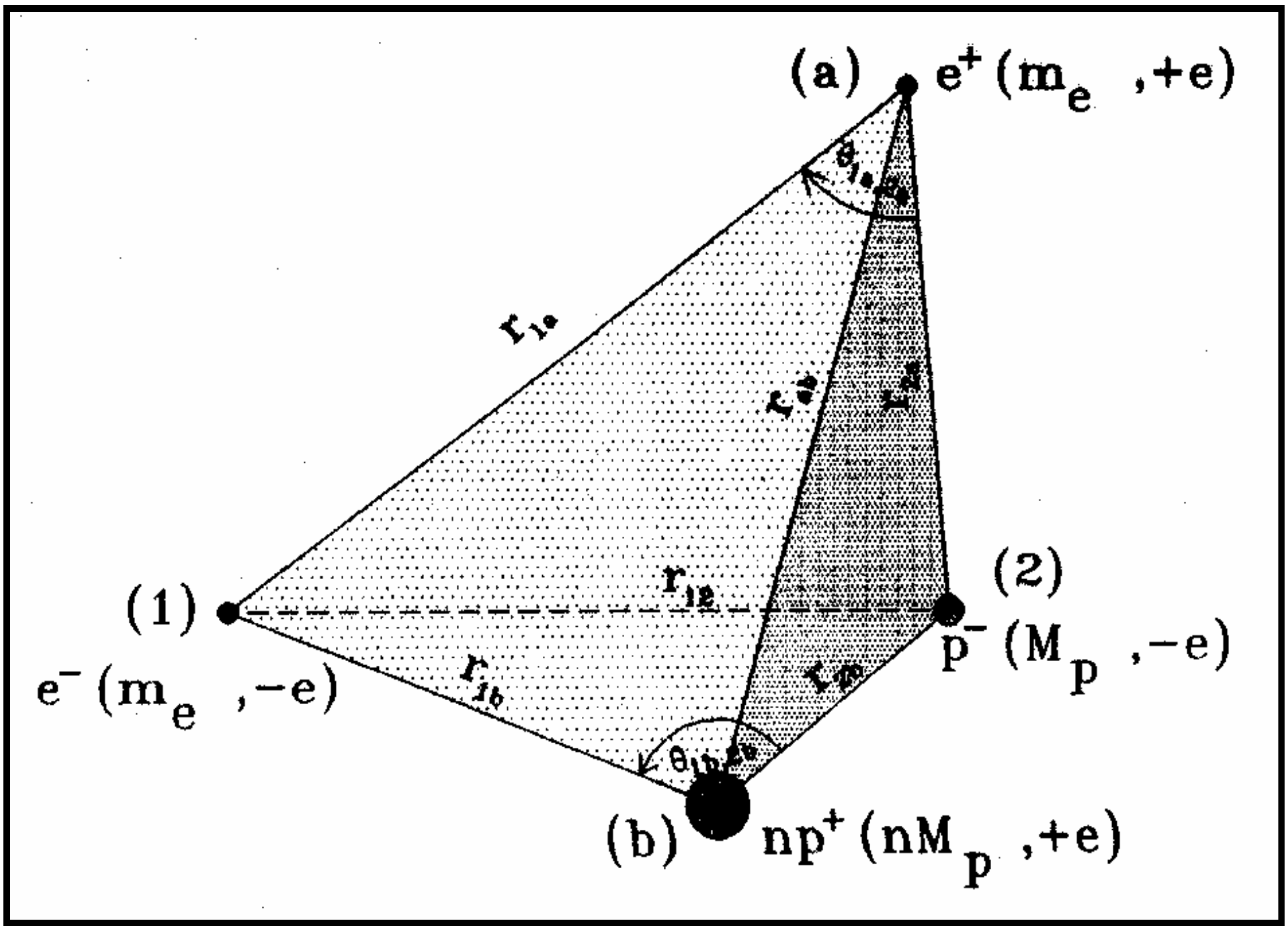

**Fig.1: Relative coordinates of Four-Body Exotic Molecules.**

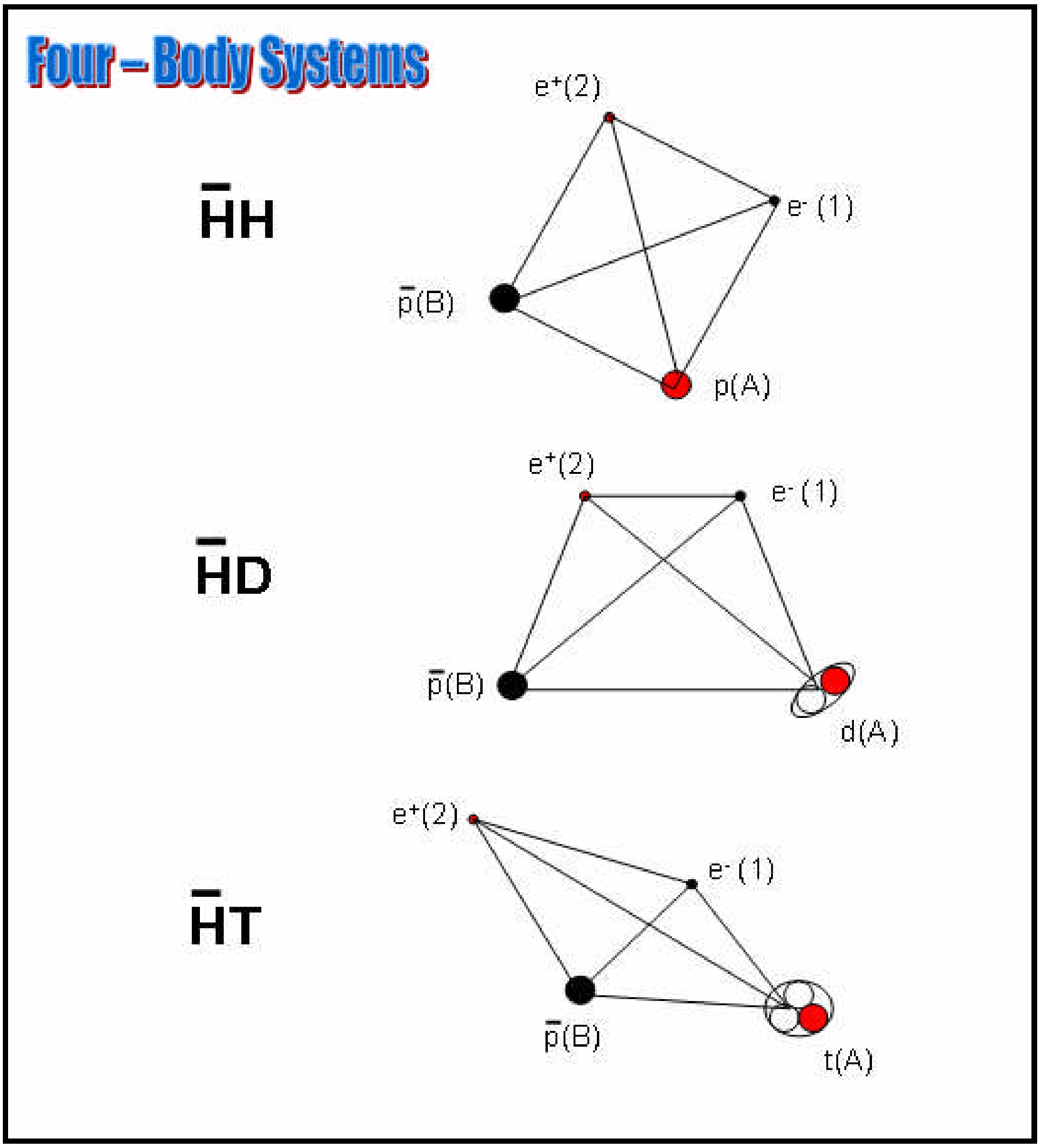

**Fig. 2: Schematic diagrams for Antihydrogen-Hydrogen, Antihydrogen-Deuterium and Antihydrogen-Tritium Exotic Molecules.**

For these systems, (see Fig. 2), the Hamiltonian can be written as

$$H = -\frac{\hbar^2}{2m_e}(\nabla_1^2 + \nabla_a^2) - \frac{\hbar^2}{2M_p}\nabla_2^2 - \frac{\hbar^2}{2nM_p}\nabla_b^2$$
$$+ Z^2 e^2 (\frac{1}{r_{12}} + \frac{1}{r_{ab}} - \frac{1}{r_{1a}} - \frac{1}{r_{2a}} - \frac{1}{r_{1b}} - \frac{1}{r_{2b}}) \quad , \tag{15}$$

On using Rydberg units (where $m_r = \frac{1}{2}$, $e^2 = 2$, $\hbar^2 = 1$, $(m_r e^4)/(2\hbar^2)$ is the unit of energy and $(\hbar^2/ m_r e^2)$ is the unit of distance), the Hamiltonian reduces (at Z=1) to the form:

$$\boldsymbol{H} = -(\nabla_1^2 + \nabla_a^2) - \sigma(\nabla_2^2 + \frac{1}{\boldsymbol{n}}\nabla_b^2)$$
$$+ (\frac{2}{\boldsymbol{r}_{12}} + \frac{2}{\boldsymbol{r}_{ab}} - \frac{2}{\boldsymbol{r}_{1a}} - \frac{2}{\boldsymbol{r}_{2a}} - \frac{2}{\boldsymbol{r}_{1b}} - \frac{2}{\boldsymbol{r}_{2b}}). \tag{16}$$

where $\sigma = m_e/M_P$ and n`=2 and n`=3 stand for $\overline{H}D$ and $\overline{H}T$ molecules, respectively.

The binding energy $W_1(\sigma)$ for such molecules will be given by;

$$W_1(\sigma) = E_1(\sigma) + 2 - \Delta \quad , \tag{17}$$

where

$$\Delta = \frac{(n-1)\sigma}{2(n\sigma+1)} \quad . \tag{18}$$

It is obvious that $\Delta$ is always positive definite, so if $W(\sigma)$ is negative for n` = 1, it is necessarily negative for all n` > 1. This means that if the four-body molecule 12ab, (see Fig. 1), exists at n` = 1 for a given $\sigma$, it exists at all values of n` for the same $\sigma$. This in turn means that the existence of $\overline{H}H$ molecule implies the existence of $\overline{H}D$ and $\overline{H}T$ molecules. It has been also concluded that:

If $W_k = E_k - E_k^{a2} - E_k^{1b}$ where $E_k^{a2}$ and $E_k^{1b}$ are both the k-th excited states of the atomic clusters *a*-2 and 1-b, respectively, and $E_k$ is the corresponding singlet excited state of the quasimolecule, the following statements are true:

(i) If the k-th excited state of 12ab exists at $\sigma = 1$, n`=1 , then the k-th excited state of 12ab exists also at $0 \le \sigma \le 1$ for all $n \ge 1$.

(ii) $W_k(0) \le W_k(\sigma) \le W_k(1)$. (19)

The proof of i follows from the definition of $W_k$ and the generalization of the theorem for $k > 1$, while (ii) is a result of (i).

## 3.1. Kinetic and Potential Energy Operators of $\overline{H}H$

Our system consists of Antihydrogen and Hydrogen atoms bound together to form a quazimolecular structure denoted by $\overline{H}H$. We will deal with this molecule as a four-particle system. A schematic diagram for the system is shown in **Fig. (2)** with n` = 1

The Hamiltonian of the $\overline{H}H$ is identical with that of eq. (16) and can be written as

$$H = [T_1 + T_a + \sigma(T_2 + T_b)]_T + V \quad , \tag{20}$$

where

$$T_i = -\nabla_i^2 \quad ; \qquad i = 1, a, 2, b \quad . \tag{21}$$

Remembering that we will not use Born-Oppenheimer approximation, instead, we will let the four particles move in space, and write the Hamiltonian in terms of the relative coordinates of the system. Accordingly, the kinetic energy operator will be reduced to the forms given at eqs. (20) and (21), where

$$\nabla_i^2 = \sum_{j\neq i}\left(\frac{\partial^2}{\partial r_{ij}^2} + \frac{2}{r_{ij}}\frac{\partial}{\partial r_{ij}}\right) + 2\sum_{j\neq i}\sum_{k\neq i}\cos\theta_{ij,ik}\frac{\partial^2}{\partial r_{ij}\partial r_{ik}} \quad . \qquad (22)$$

$$j < k$$

and

$$\cos\theta_{ij,ik} = \frac{r_{ij}^2 + r_{ik}^2 - r_{jk}^2}{2\, r_{ij}\, r_{ik}} \quad . \qquad (23)$$

## 3.2. Hylleraas' Coordinates.

Our system has to be described in a more suitable set of coordinates; this is because the components of the spherical polar coordinates $r_{ij}$ are not orthogonal to each other and so highly dependent. Besides, the new set has to take in consideration the correlation relation between different particles in the molecule. We notice that the interaction between different particles in our system is one-to-two instead two-body interaction.

The most suitable set of coordinates for describing our molecules is composed of elliptic coordinates, some times also known as Hylleraas' coordinates. It consists of confocal ellipses and hyperbolas and is defined by

$$s_i = (r_{ia} + r_{ib})/r_{ab} \qquad ; \qquad i = 1, 2 \tag{24a}$$

$$t_i = (r_{ia} - r_{ib})/r_{ab} \qquad ; \qquad i = 1, 2 \tag{24b}$$

$$s_a = (r_{1a} + r_{2a})/r_{12} \qquad , \qquad s_b = (r_{1b} + r_{2b})/r_{12}\,, \tag{24c}$$

$$t_a = (r_{1a} - r_{2a})/r_{12} \qquad , \qquad t_b = (r_{1b} - r_{2b})/r_{12}\,, \tag{24d}$$

$$u = r_{12}/r_{ab} \qquad , \qquad v = r_{ab}\,. \tag{24e}$$

As it is evident $s_i$, $s_a$ and $s_b$ are constants on ellipses the distance between their two foci are $r_{ab}$ for $s_i$ and $r_{12}$ for $s_a$ and $s_b$, while $t_i$, $t_a$ and $t_b$ are constants on hyperbolas again the distances between their two foci are $r_{ab}$ for $t_i$ and $r_{12}$ for $t_a$ and $t_b$. It is clear that s goes from 1 to infinity, t goes from $-1$ to 1, and v goes from 0 to infinity. In addition to these variables there are also the angles of rotation $\Phi$'s about the axis joining the two foci.

The point now is to write the Hamiltonian of the system in terms of these coordinates. The partial derivatives with respect to $r_{ia}$ and $r_{ib}$ can be expressed in terms of $s_i$ and $t_i$ as follows;

$$\frac{\partial}{\partial r_{1a}} = \frac{\partial s_1}{\partial r_{1a}}\frac{\partial}{\partial s_1} + \frac{\partial t_1}{\partial r_{1a}}\frac{\partial}{\partial t_1} \quad , \tag{25}$$

then

$$\frac{\partial}{\partial r_{1a}} = \frac{1}{v}\left(\frac{\partial}{\partial s_1} + \frac{\partial}{\partial t_1}\right) \quad , \quad \frac{\partial^2}{\partial r_{1a}^2} = \frac{1}{v^2}\left(\frac{\partial^2}{\partial s_1^2} + 2\frac{\partial^2}{\partial s_1 \partial t_1} + \frac{\partial^2}{\partial t_1^2}\right), \tag{26a}$$

and so on. Finally we have

$$\frac{\partial}{\partial r_{12}} = \frac{1}{v}\frac{\partial}{\partial u} \qquad , \qquad \frac{\partial^2}{\partial r_{12}^2} = \frac{1}{v^2}\frac{\partial^2}{\partial u^2} \quad . \tag{26b}$$

We also have

$$\cos\theta_{ij,ik} = \frac{r_{ij}^2 + r_{ik}^2 - r_{jk}^2}{2\, r_{ij}\, r_{ik}} \qquad , \qquad i = 1, 2$$

$$= \frac{\frac{v^2}{4}[(s_i + t_i)^2 + (s_i - t_i)^2] - v^2}{\frac{2v^2}{4}(s_i + t_i)(s_i - t_i)} \quad ,$$

$$\therefore \qquad \cos\theta_{ia,ib} = \frac{s_i^2 + t_i^2 - 2}{s_i^2 - t_i^2} \quad . \qquad (27)$$

Now, the kinetic energy operator (for example) T1 can be written as

$$T_1 = -\frac{4}{v^2}\frac{1}{(s_1^2 - t_1^2)}[(s_1^2 - 1)\frac{\partial^2}{\partial s_1^2} + (1 - t_1^2)\frac{\partial^2}{\partial t_1^2} + 2s_1\frac{\partial}{\partial s_1}$$

$$-2t_1\frac{\partial}{\partial t_1}] - \frac{1}{v^2}[\frac{\partial}{\partial u} + \frac{2}{u} + 2(\cos\theta_{12,1a} + \cos\theta_{12,1b})\frac{\partial}{\partial s_1}$$

$$+2(\cos\theta_{12,1a} - \cos\theta_{12,1b})\frac{\partial}{\partial t_1}]\frac{\partial}{\partial u} \quad ,$$

(28)

Similar forms can be derived for $T_2$, $T_a$ and $T_b$.

On the other hand, the potential energy operator which is given by;

$$V = 2(\frac{1}{r_{12}} + \frac{1}{r_{ab}} - \frac{1}{r_{1a}} - \frac{1}{r_{2a}} - \frac{1}{r_{1b}} - \frac{1}{r_{2b}}),$$

will be modified to;

$$V = \frac{2}{v}[(\frac{1}{u} + 1) - \frac{4s_1}{(s_1^2 - t_1^2)} - \frac{4s_2}{(s_2^2 - t_2^2)}] \quad . \qquad (29)$$

**3.3. The Volume Element of the System.**

Two forms for the volume element $d\tau$ should be distinguished:

(i) v is variable

In this case the volume element takes the form

$$d\tau = \frac{v^8}{64}(s_1^2 - t_1^2)(s_2^2 - t_2^2)ds_1 ds_2 dt_1 dt_2 dv$$

$$\sin\theta_v d\theta_v d\varphi_v d\Phi_1 d\Phi_2 \quad . \qquad (30)$$

with

$$1 \le s_i \le \infty \quad , \quad -1 \le t_i \le 1 \quad , \quad 0 \le v \le \infty \quad , \qquad (31a)$$

$$0 \le \theta_v \le \pi \quad , \quad 0 \le \varphi_v \le 2\pi \quad , \quad 0 \le \Phi_i \le 2\pi \quad . \qquad (31b)$$

(ii) $r_{ab}$ is Constant:

In this case, the volume element is reduced to

$$d\tau = \frac{v^6}{64}(s_1^2 - t_1^2)(s_2^2 - t_2^2)\, ds_1 ds_2 dt_1 dt_2 d\Phi_1 d\Phi_2 . \qquad (32)$$

## 3.4. Wavefunctions of $\overline{\mathrm{H}}\mathrm{H}$

In constructing the wavefunction of the first heterohydrogen molecule, (see Fig.3) one has to take into account that the system consists of four bodies: two particles, $e^-$ and $p^+$, and two antiparticles, $e^+$ and $p^-$. The two foci of the system will be taken as $p^+$ and $e^+$ with a distance v from each other, due to the elliptic coordinates. Such arrangement for the molecule is considered because we are dealing with particle-antiparticle system in which the two p's, $p^+$ and $p^-$, have the same mass and opposite charge and so tend to attract each other leading to the annihilation of the molecule. Besides, such arrangement will justify the important symmetry property, between $e^+$ and $p^+$ from one side and $e^-$ and $p^-$ from the other side, for the wavefunction.A suitable molecular wavefunction containing the full number of the required coordinates should be built. For this reason, the wavefunction has to include two main parts: one describes the ability of different particles in the molecule to be bound and the other describes the dissociation of the molecule into separate atoms. Besides, it will also contain powers of coordinates to describe possible deformation in the molecular orbitals in the high quantum states of the molecule.

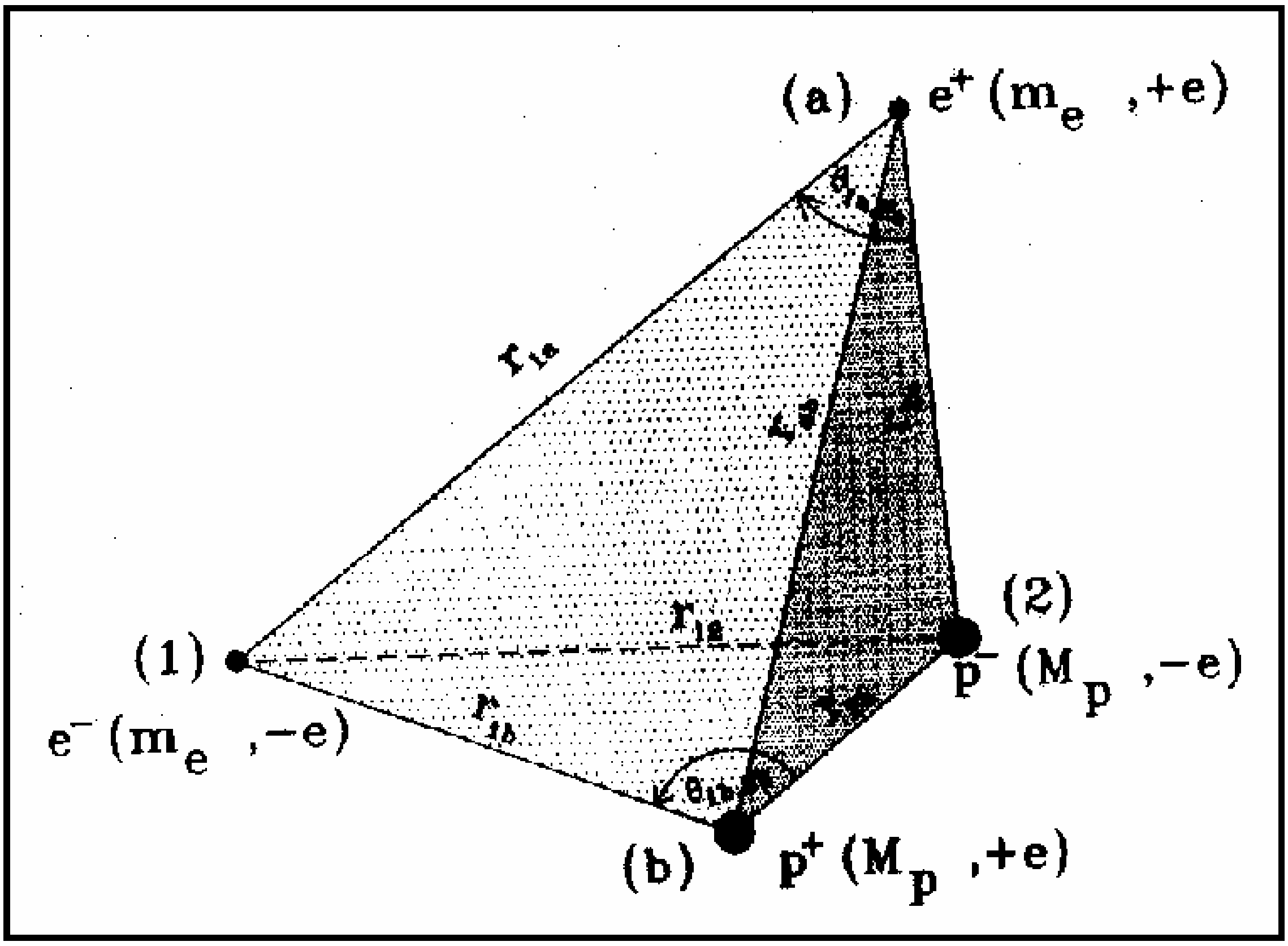

**Fig. 3: Relative coordinates of Antihydrogen-Hydrogen Exotic Molecules.**

Now, having such imagination about the physics that the wavefunction must satisfy, we can put a form for it. Our trial j-th component of the wavefunction will take the form;

$$\left|\chi_j\right\rangle = s_1^{m_j}\, s_2^{n_j}\, e^{-\alpha_j (s_1+s_2)}\, t_1^{k_j}\, t_2^{\ell_j} \cosh\left[\beta_j (t_1 - t_2)\right] u^{p_j}\, v^{q_j}\, e^{-\gamma_j v} \quad . \quad (33)$$

The total wavefunction for the system is then a superposition of the different molecular states where each state will be given by;

$$\left|\psi_k\right\rangle = \sum_{j=0}^{\infty} c_{kj} \left|\chi_j\right\rangle \quad , \tag{34}$$

where k stands for the state of the system; $k = 1$ stands for the ground-state, $k = 2$ is the first excited state … etc.

Applying now the kinetic energy operators given at eqs. (28) to the j-th component $\chi_j$ of the wavefunction, eq. (33), we obtain the following relations:

$$\begin{aligned}
T_1 \chi_j = \frac{1}{v^2 (s_1^2 - t_1^2)} \Big\{ & [-4 m_j (m_j - 1) + 4 k_j (k_j - 1) - 4 (p_j + 2)(m_j - k_j) \\
& + 4(\alpha_j^2 - \beta_j^2)] + [4 m_j (m_j - 1)] s_1^{-2} - [4 k_j (k_j - 1)] t_1^{-2} - (8 m_j \alpha_j) s_1^{-1} \\
& - (8 k_j \beta_j) t_1^{-1} \tanh[\beta_j (t_1 - t_2)] + [4\alpha_j (2 m_j + 2 + p_j)] s_1 + [4\beta_j (2 k_j \\
& + 2 + p_j)] t_1 \tanh[\beta_j (t_1 - t_2)] - (4\alpha_j^2) s_1^2 + (4\beta_j^2) t_1^2 + (p_j \alpha_j) s_1^3 u^{-2} \\
& + (p_j \beta_j) t_1^3 u^{-2} \tanh[\beta_j (t_1 - t_2)] - [p_j (m_j + k_j + p_j + 1)] s_1^2 u^{-2} \\
& + [p_j (m_j + k_j + p_j + 1)] t_1^2 u^{-2} - (p_j \alpha_j) s_1 t_1^2 u^{-2} - (p_j \beta_j) t_1 s_1^2 u^{-2} \\
& \times \tanh[\beta_j (t_1 - t_2)] + [p_j (m_j - k_j)] s_2^2 u^{-2} + [p_j (m_j - k_j)] t_2^2 u^{-2} \\
& - (p_j \alpha_j) s_1 s_2^2 u^{-2} - (p_j \alpha_j) s_1 t_2^2 u^{-2} - (p_j \beta_j) t_1 s_2^2 u^{-2} \tanh[\beta_j (t_1 - t_2)] \\
& - (p_j \beta_j) t_1 t_2^2 u^{-2} \tanh[\beta_j (t_1 - t_2)] + (2 p_j \beta_j) s_1 s_2 t_2 u^{-2} \\
& \times \tanh[\beta_j (t_1 - t_2)] + (2 p_j k_j) s_1 s_2 t_2 t_1^{-1} u^{-2} + (2 p_j \alpha_j) t_1 s_2 t_2 u^{-2} \\
& - (2 p_j m_j) t_1 s_2 t_2 s_1^{-1} u^{-2} \Big\} \chi_j \, ,
\end{aligned} \tag{35a}$$

$$T_2\,\chi_j = \frac{1}{v^2(s_2^2 - t_2^2)}\Big\{[-4n_j(n_j-1) + 4\ell_j(\ell_j-1) - 4(p_j+2)(n_j-\ell_j)$$

$$+ 4(\alpha_j^2 - \beta_j^2)] + [4n_j(n_j-1)]s_2^{-2} - [4\ell_j(\ell_j-1)]t_2^{-2} - (8n_j\alpha_j)s_2^{-1}$$

$$+ (8\ell_j\beta_j)t_2^{-1}\tanh[\beta_j(t_1-t_2)] + [4\alpha_j(2n_j+2+p_j)]s_2 - [4\beta_j(2\ell_j$$

$$+ 2 + p_j)]t_2\tanh[\beta_j(t_1-t_2)] - (4\alpha_j^2)s_2^2 + (4\beta_j^2)t_2^2 + (p_j\alpha_j)s_2^3u^{-2} -$$

$$- (p_j\beta_j)t_2^3u^{-2}\tanh[\beta_j(t_1-t_2)] - [p_j(n_j+\ell_j+p_j+1)]s_2^2u^{-2}$$

$$+ [p_j(n_j+\ell_j+p_j+1)]t_2^2u^{-2} - (p_j\alpha_j)s_2t_2^2u^{-2} + (p_j\beta_j)t_2s_2^2u^{-2}$$

$$\times\tanh[\beta_j(t_1-t_2)] + [p_j(n_j-\ell_j)]s_1^2u^{-2} + [p_j(n_j-\ell_j)]t_1^2u^{-2}$$

$$- (p_j\alpha_j)s_2s_1^2u^{-2} - (p_j\alpha_j)s_2t_1^2u^{-2} + (p_j\beta_j)t_2s_1^2u^{-2}\tanh[\beta_j(t_1-t_2)]$$

$$+ (p_j\beta_j)t_2t_1^2u^{-2}\tanh[\beta_j(t_1-t_2)] - (2p_j\beta_j)s_2s_1t_1u^{-2}$$

$$\times\tanh[\beta_j(t_1-t_2)] + (2p_j\ell_j)s_2s_1t_1t_2^{-1}u^{-2} + (2p_j\alpha_j)t_2s_1t_1u^{-2} \qquad (35b)$$

$$- (2p_jn_j)t_2s_1t_1s_2^{-1}u^{-2}\Big\}\chi_j\ .$$

Now, we deal with our system through the coordinates $s_1, s_2, t_1, t_2$, $u, v$ rather than $s_a, s_b, t_a, t_b, u, v$. We notice from eqs.(20)-(23) that $T_a$ and $T_b$ are obtained from $T_1$ and $T_2$, respectively, by replacing $s_1, s_2, t_1, t_2, u, v$ by $s_a, s_b, t_a, t_b, u, v$, respectively. Thus, $T_a\,\chi_j$ and $T_b\,\chi_j$ can be derived via (35a) and (35b), respectively, using the same rearrangement.

Finally, from (29), applying the potential energy operator to the j-th component of the wavefunction, eq. (33), we find

$$V\chi_j = \frac{2}{v}\Big[\Big(\frac{1}{u} + 1\Big) - \frac{4s_1}{s_1^2 - t_1^2} - \frac{4s_2}{s_2^2 - t_2^2}\Big]\chi_j\ . \qquad (36)$$

The matrix elements required for RRVM or VT possess very complicated forms and will not be presented here explicitly.

## 4. RESULTS AND DISCUSSIONS

The computational part of the work has one main goal, namely to test the possible existence and formation of heterohydrogen molecules ($\bar{H}H$, $\bar{H}D$ and $\bar{H}T$).The first step in the calculations is to optimize the parameters $\alpha$, $\beta$, and $\gamma$ involved in the wavefunction eq. (33) with respect to the energy (see e.g. Figs.4 and 5).

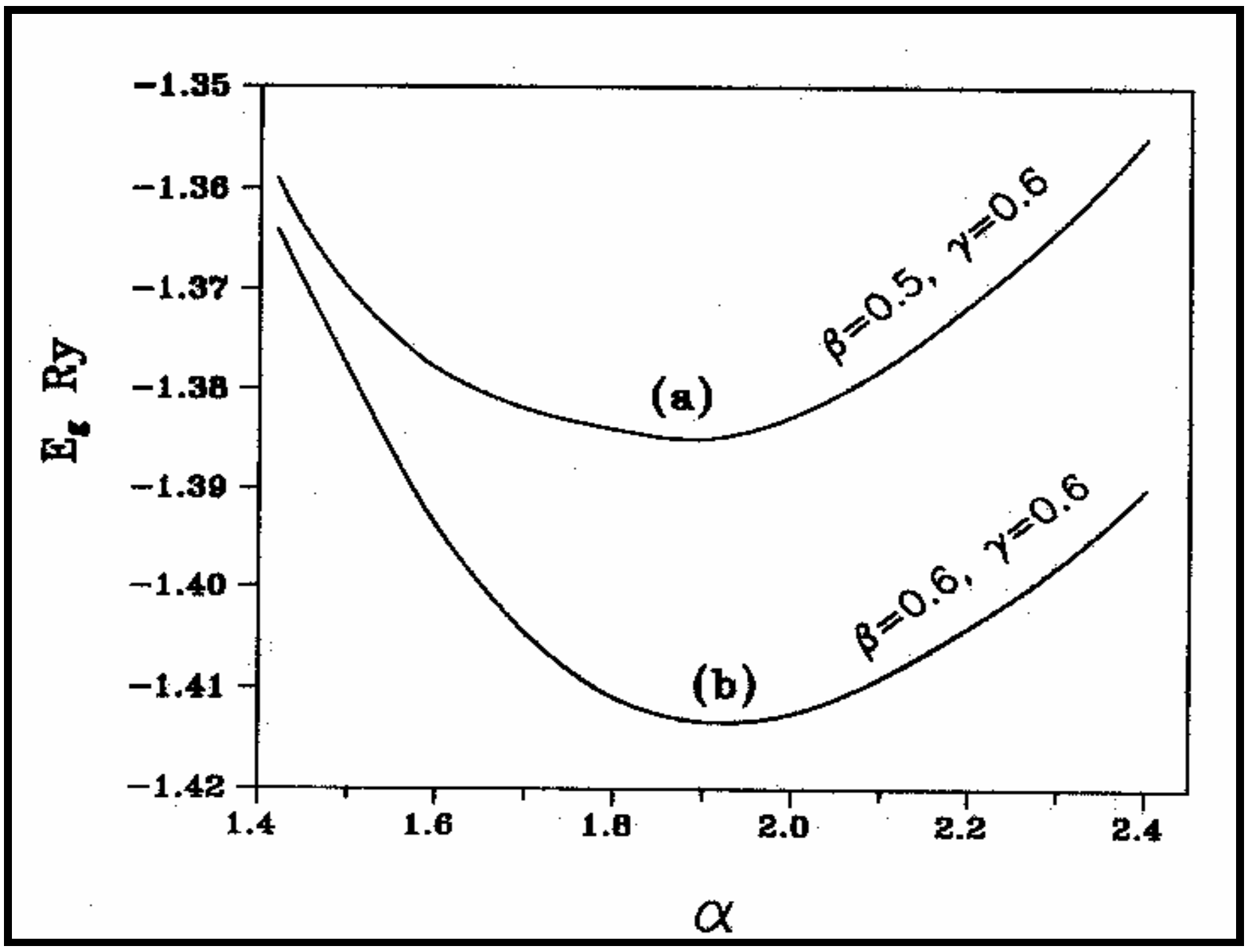

**Fig4: Variation of the ground state energy of $\bar{H}$H with $\alpha$ at constant value of $\gamma$ and two values of $\beta$. Note that $\alpha$ =1.9.0 gives the minimum energy $\alpha$ ~ 1.9 gives the minimum energy**

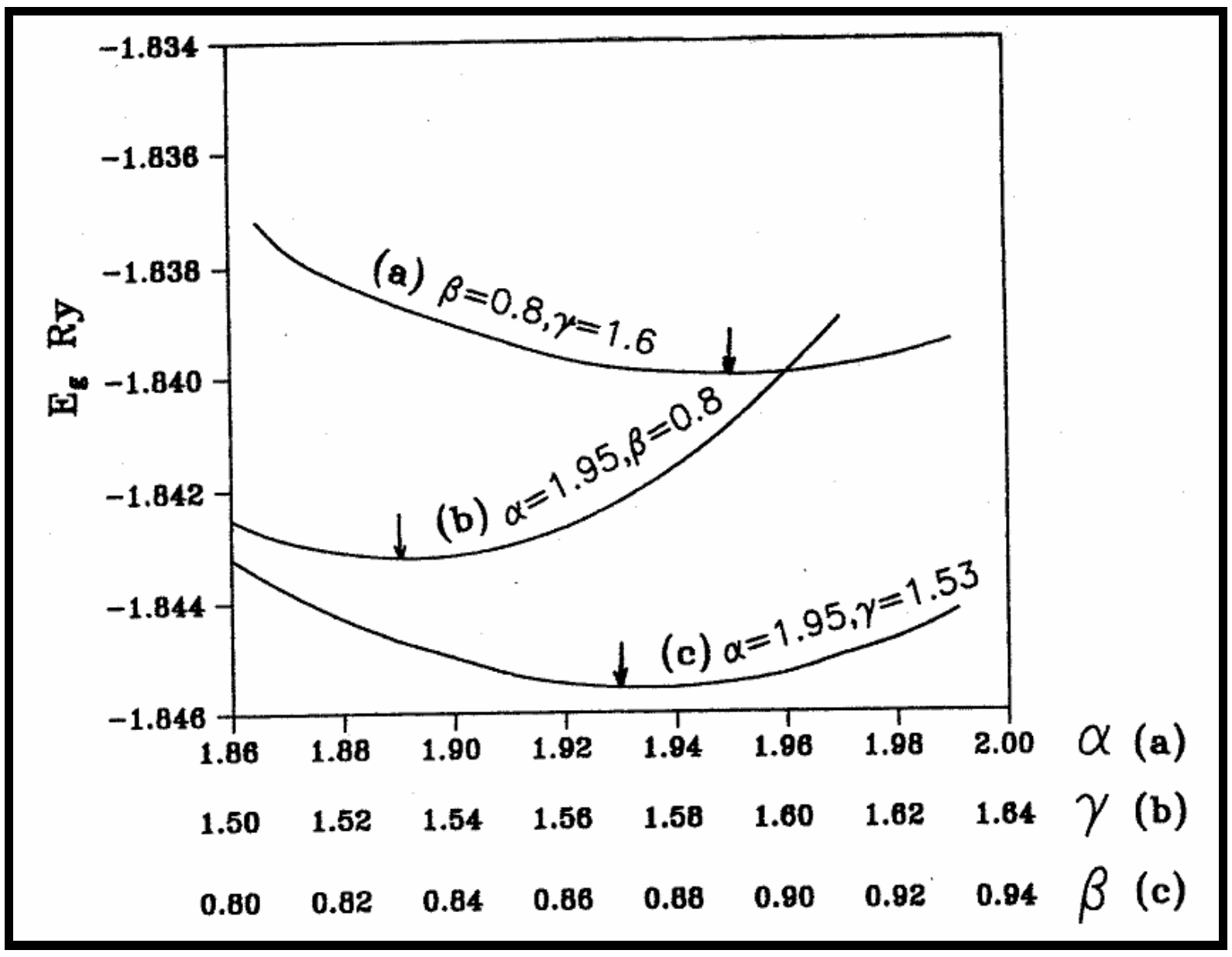

**Fig.5: Minimization of the ground-state energy of $\bar{H}H$ with respect to the nonlinear parameters α, γ and β. The arrows show the optimized values of the parameters.**

Many investigations have been carried out in this direction and provided us with the best values α = 1.95, β = 0.87 and γ = 1.53 when the first five components of the wavefunction (see Table 1) are considered. With these values, the convergence of the total and binding energies of the ground-state of $\bar{H}H$ molecule has been studied when n is steadily increasing (see Table 2). The Table shows that with only 13 components of the wavefunction, the molecule is bound, and its binding energy can be lowered using 25 components to - 0.7476 eV. This result is in complete agreement with the extension of the theorem of four-body systems and ends, for the first time, the dispute about the existence of the $\bar{H}H$ as a molecule composed of an antihydrogen and a hydrogen atom.

**Table (1): Powers of the elements of the wavefunction (eq. 33)**

| j | $m_j$ | $n_j$ | $k_j$ | $\ell_j$ | $q_j$ | $p_j$ |
|---|---|---|---|---|---|---|
| 1 | 0 | 0 | 0 | 0 | 0 | 0 |
| 2 | 0 | 0 | 0 | 0 | 1 | 0 |
| 3 | 0 | 0 | 0 | 0 | 2 | 0 |
| 4 | 0 | 0 | 1 | 1 | 0 | 0 |
| 5 | 0 | 0 | 1 | 1 | 1 | 0 |
| 6 | 0 | 0 | 2 | 2 | 0 | 0 |
| 7 | 0 | 0 | 2 | 2 | 1 | 0 |
| 8 | 1 | 1 | 0 | 0 | 0 | 0 |
| 9 | 1 | 1 | 0 | 0 | 1 | 0 |
| 10 | 1 | 1 | 1 | 1 | 0 | 0 |
| 11 | 1 | 1 | 1 | 1 | 1 | 0 |
| 12 | 1 | 1 | 2 | 2 | 0 | 0 |
| 13 | 1 | 1 | 2 | 2 | 1 | 0 |
| 14 | 0 | 0 | 0 | 0 | 3 | 0 |
| 15 | 0 | 0 | 1 | 1 | 2 | 0 |
| 16 | 0 | 0 | 2 | 2 | 2 | 0 |
| 17 | 1 | 1 | 0 | 0 | 2 | 0 |
| 18 | 1 | 1 | 1 | 1 | 2 | 0 |
| 19 | 1 | 1 | 2 | 2 | 2 | 0 |
| 20 | 0 | 0 | 1 | 1 | 3 | 0 |
| 21 | 0 | 0 | 2 | 2 | 3 | 0 |
| 22 | 1 | 1 | 0 | 0 | 3 | 0 |
| 23 | 1 | 1 | 1 | 1 | 3 | 0 |
| 24 | 1 | 1 | 2` | 2 | 3 | 0 |
| 25 | 0 | 0 | 0 | 0 | 4 | 0 |
| 25 | 0 | 0 | 0 | 0 | 0 | 1 |

It is interesting to mention that all attempts we made to test the stability of this molecule at higher states have failed.

**Table 2: Ground state Energy $E_g$ and Binding Energy $W_g$ of $\bar{H}H$ using up to 25 terms of the wavefunction (33)**

| No. of terms | $E_g$ (Ry) | $W_g$ (eV) |
|---|---|---|
| 1 | -1.4727 | - |
| 5 | -1.8455 | - |
| 9 | -1.8938 | - |
| 13 | -2.0114 | -0.1552 |
| 19 | -2.0301 | -0.4096 |
| 23 | -2.05429 | -0.7384 |
| 25 | -2.05497 | -0.7476 |

Calculations of the binding energies of $\bar{H}D$ and $\bar{H}T$ using the same values of the nonlinear parameters employed for $\bar{H}H$ (i.e. $\alpha = 1.95, \beta = 0.87$ and $\gamma = 1.53$) yield - 0.7507 eV and – 0.7521 eV as ground state energies for $\bar{H}D$ and HT, respectively. This means that the possible existence and formation of these molecules in nature have been confirmed. Again, no excited states of these molecules have shown up.

Further investigations of the effect of adding components depending on u to the set of 25 basis functions have shown that a single component could lower considerably the binding energies of heterohydrogens to -0.9325 , -0.9357 and -0.9373 eV (for n`= 1,2,3, respectively). (This large contribution is attributed to large number of components of the trial wavefunctions (depending on $s_1$, $s_2$, $t_1$, $t_2$, etc.) covered by this term only and added to previous 25 components; a matter which is connected with drastic increase in the computer time of the variational energies).

The same basis set leads, within the framework of the virial theory, to -0.9335, -0.9370 and -0.9383 eV, as the binding energies of antihydrogen-hydrogen, antihydrogen-deuterium and antihydrogen-tritium molecules, respectively. This supports the argument that the VT leads to slight correction of the binding energies when they are close to convergence.

The expectation values of the internal distances of the three heterohydrogens have been calculated using the lastly mentioned forms of the wavefunctions (i.e. 25 components + u). The results are displayed in Table 3.

**Table 3: Expectation values of < $r_{ij}$ > measured in Bohr's radius $a_o$ for different heteromolecules**

| $r_{ij}$ | $\bar{H}$**H** | $\bar{H}$**D** | $\bar{H}$**T** |
|---|---|---|---|
| $r_{ab}=r_{12}$ | **2.71168** | **2.7114** | **2.71135** |
| $r_{1b}=r_{2b}=r_{1a}=r_{2a}$ | **2.18440** | **2.1842** | **2.18410** |

**Table 4: Binding Energies of Antihydrogen exotic Molecules**

| | |
|---|---|
| Antihydrogen-Hydrogen | - 0.7476 eV |
| Antihydrogen-Deuterium | - 0.7507 eV |
| Antihydrogen – Tritium | -0.7521 eV |

It has been found that $<r_{ab}> = <r_{12}>$ and $<r_{1b}> = <r_{2a}> = <r_{1a}> = <r_{2b}>$ for all molecules. Table 3 confirms the idea that the size of the heteromolecule gets smaller the larger is the mass of the nucleus of the involved atom A final remark could be made on the lengths of the internal distances in comparison with nucleus-nucleus distance in the hydrogen molecule (about 1.4 a.u) which might support the argument that heterohydrogens are larger in size than the hydrogen molecules and the light particles play an important role in the screening of heavy nucleons, a matter which reduces the possibility of dissociation into protoniums and positroniums as well as their formation as intermediate states leading to annihilation.

## 5. CONCLUSIONS

The most interesting conclusions of the present work can be summarized in the following points:

a) The stability of heterohydrogens as molecular structures against dissociation into antihydrogen and hydrogen, deuterium or tritium has been computationally confirmed for the first time.

b) The binding energy and size of a heterohydrogen are smaller the heavier is the nucleus of involved atom.

c) Both conclusions a) and b) agree completely with the theorem of four-body molecules ([26], [32] and [33]).

d) On the other hand, having in mind the adiabatic picture of the systems considered, according to which their binding energies should lie considerably lower than the calculated ones, our results suggest the possible formation of these molecules as resonant states in Antihydrogen-Atom interaction.

e) The results presented in the preceding points should enhance further computational and experimental investigations of the present systems.

.